# THE FUTURE PROSPECTS OF MUON COLLIDERS AND NEUTRINO FACTORIES

MANUELA BOSCOLO*

*Istituto Nazionale Fisica Nucleare, Laboratori Nazionali di Frascati*
*Via E. Fermi 40, 00444 Frascati (Rome), Italy*
manuela.boscolo@lnf.infn.it

JEAN-PIERRE DELAHAYE

*CERN, 1211 Geneva 23, Switzerland*
jean-pierre.delahaye@cern.ch

MARK PALMER

*Brookhaven National Laboratory*
*Upton, New York 11973, USA*
mpalmer@bnl.gov

The potential of muon beams for high energy physics applications is described along with the challenges of producing high quality muon beams. Two proposed approaches for delivering high intensity muon beams, a proton driver source and a positron driver source, are described and compared. The proton driver concepts are based on the studies from the Muon Accelerator Program (MAP). The MAP effort focused on a path to deliver muon-based facilities, ranging from neutrino factories to muon colliders, that could span research needs at both the intensity and energy frontiers. The Low EMittance Muon Accelerator (LEMMA) concept, which uses a positron-driven source, provides an attractive path to very high energy lepton colliders with improved particle backgrounds. The recent study of a 14 TeV muon collider in the LHC tunnel, which could leverage the existing CERN injectors and infrastructure and provide physics reach comparable to the 100 TeV FCC-hh, at lower cost and with cleaner physics conditions, is also discussed. The present status of the design and R&D efforts towards each of these sources is described. A summary of important R&D required to establish a facility path for each concept is also presented.

## 1. Introduction

The major 2012 discoveries of the large flavor mixing angle $\theta_{13}$ at Daya Bay in China and of the Higgs boson by the LHC at CERN dramatically modified the Particle Physics landscape. Although the Higgs discovery provides a splendid confirmation of the Standard Model (SM), no sign of physics Beyond the Standard Model (BSM) has yet been detected at the LHC. BSM physics is necessary to answer key questions that the SM does not address – in particular, dark matter, dark energy, the matter-antimatter asymmetry, and the mass of the neutrino. Therefore, the quest for BSM physics is a high priority for the future of High Energy Physics (HEP).

In order to probe BSM physics, HEP requires capabilities at both the high energy and high intensity frontiers. Neutrino oscillations are irrefutable evidence for BSM physics that have the potential to probe extremely high energy scales. While the large value of the flavor mixing angle means that the Long Baseline Neutrino Facility (LBNF) at Fermilab and the Deep Underground Neutrino Experiment (DUNE) at the Sanford Underground Research Facility (SURF) can provide the performance needed to carry out the next

---

manuela.boscolo@lnf.infn.it

generation of neutrino mixing experiments, a Neutrino Factory (NF), with its intense and well-defined flux of neutrinos from muon decay, is the tool best-suited to support a program in precision flavor physics at the intensity frontier. At the energy frontier, the energy scale of new BSM physics has moved into the TeV regime. Thus, a multi-TeV lepton collider, possibly a Muon Collider (MC), can provide a precision facility to complement the LHC when evidence for such new physics is confirmed.

The remainder of this article is organized as follows:
- Section 1 – Introduction and potential physics reach of muon beams.
- Section 2 – An overview of the muon production schemes using a proton-based source and a positron-based source. For each source, a summary of the key design elements followed by an overview of the R&D status is presented.
- Section 3 – An overview of the NF and MC applications that can be supported by each source technology.
- Section 4 – The authors' perspective on the future R&D required to achieve a muon accelerator capability to support the science needs of the high energy physics community.
- Section 5 – Concluding remarks.

## 1.1. *The Beauty and Challenge of Muon Beams*

The development of Muon Colliders has been extensively reviewed in a previous *Reviews of Accelerator Science and Technology* article [1]. The present review updates the progress on their design and describes the substantial R&D progress towards establishing their feasibility, that has been achieved in the interim. It outlines the key concepts that have emerged that offer significant potential improvements to MC performance.

As pointed out in Ref. [1], the advantage of muons is that they are fundamental leptons with a mass that is a factor of 207 greater than that of their lighter companions, the electron and positron. Thus, the power emitted by a muon beam as synchrotron radiation, which scales as the fourth power of the particle mass, is reduced by nine orders of magnitude relative to that emitted by electrons having the same energy and bending radius.

In order to produce collisions with significant luminosity at very high energies, electron bunches must be accelerated in linear colliders and focused to extremely small dimensions to collide only once in a single detector. In contrast, muon bunches can be accelerated in multi-pass rings and then brought into collisions for multiple passes in collider rings equipped with multiple detectors. The ability to utilize multi-pass rings for both muon acceleration and luminosity production results in Muon Colliders that are more efficient and more compact than electron-positron linear colliders, thus opening up the possibility of very significant power and cost savings. At TeV-scale energies, very high luminosities can be achieved in a muon collider with beam emittances far larger than those required for the corresponding electron machines. Furthermore, the large cross section for s-channel resonances from $\mu^+\mu^-$ versus $e^+e^-$ collisions means that a MC can provide excellent sensitivity for key physics studies.

As shown in Fig. 1 [1, 2], which compares a figure of merit, defined as the luminosity normalized to the wall plug power required by the facility, a MC with a center-of-mass (CoM) energy greater than ~2 TeV compares quite favorably with all other lepton technologies. A MC at the TeV-scale can provide higher luminosity with significantly lower overall power consumption than any linear collider technology operating at the same CoM energy, even when one assumes the use of advanced concepts such as plasma wakefield acceleration technology.

When one also includes the impacts of beamstrahlung-driven spread in the effective CoM collision energy, the luminosity that is produced with an effective energy within 1% of the nominal energy at the interaction point is not affected for the MC whereas it is significantly reduced for $e^+e^-$ collisions. Thus, the MC is ideal to search for new physics based on missing energy topologies, and for resolving new narrow resonances. It offers the greatest potential for physics with leptons at the energy frontier in the multi-TeV energy range.

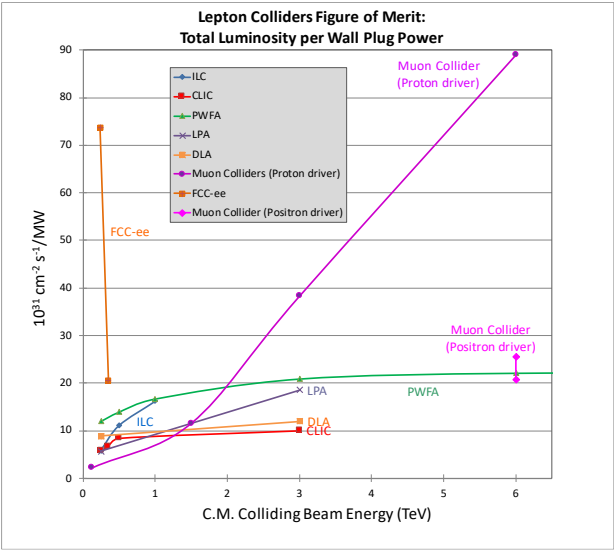

Figure 1. Figure of merit (defined as the luminosity per wall plug power) of various lepton collider technologies [1,2].

A MC is also an ideal discovery machine in the multi-TeV range. Above roughly 1 TeV, the W fusion cross section becomes dominant. Effectively, this makes the MC an electroweak boson collider with tremendous energy frontier discovery potential. It also enables extremely detailed electroweak physics studies, such as the measurement of the Higgs boson self-coupling with accuracy better than 10%.

At lower energies, the very low energy spread that can be achieved in the proton-driver based muon production and cooling scheme (a few parts in $10^5$) enables the possibility of a muon-based Higgs Factory. Such a machine would take advantage of s-channel Higgs production with muons, which is enhanced by a factor of 40,000 over what can be achieved in electron-positron collisions. With the small energy spread that can be achieved with a MC operating at ~125 GeV, such a machine offers the only path to directly scanning the Higgs resonance in order to measure its mass and width with precisions of <0.1 MeV and <0.5 MeV, respectively.

While the promise for muon beams for collider applications is clear, the short muon lifetime of 2.2 μs at rest and the difficulty of producing large numbers of muons in bunches with small emittance offer unique challenges for collider facilities. Proton-driver based sources have been pursued due to the possibility of producing relatively large numbers of muons through tertiary production from protons hitting a target. The resulting muons occupy a large phase-space and must be cooled in order to provide usable beams. In contrast, photon- and positron-based sources provide much lower production rates at lower emittance. All of these sources require rapid acceleration of the resulting muons and colliders designed to operate with decaying beams. Thus, any path to a MC requires development of several demanding technologies and innovative concepts.

At the same time, muon decays offer a window on another important area of physics. The muons, which decay predominantly according to the following processes, can provide precision sources of electron and muon neutrinos and anti-neutrinos:

$$\mu^+ \to e^+ \nu_e \bar{\nu}_\mu$$

$$\mu^- \to e^- \bar{\nu}_e \nu_\mu. \quad (1)$$

These neutrinos can serve as the source for a Neutrino Factory, which would constitute the ideal successor to the current generation of long baseline facilities that are based on more conventional technology where neutrinos are produced as secondary particles from pion decay. A NF source and detector would provide very attractive enhancements [3] for physics studies at a long baseline facility with:

- an ideal source of well-defined electron and muon neutrinos in equal quantities;
- the production via mixing of all neutrino species allowing physics with multiple channels;
- a neutrino beam composition understood with a precision of ~1%, thus providing excellent systematics;
- a clean muon detector with a magnetic field to distinguish $\mu^+$ from $\mu^-$.

### 1.2. *Enabling a Series of Facilities at the Intensity and Energy Frontiers*

Because of their great potential along with their inherent challenges, muon-based facilities have been studied for more than thirty years. The idea of Muon Colliders was first introduced in the early 1980s [4, 5] and was further developed by a series of world-wide collaborations [6] culminating in 2011 with the U.S. Muon Accelerator Program (MAP) [7]. MAP continued to develop the concepts and address the feasibility of the novel technologies required for Muon Colliders and Neutrino Factories based on a proton driver source [8, 9]. A staging scenario for a series of muon-based facilities [2] with progressively increasing complexity, where each stage can provide unique physics reach, is presented in Section 3.1. Such a staging scheme is meant to represent the potential opportunities for producing physics results, as driven by the scientific needs identified by the community, as opposed to a predetermined path for a muon-based facility. While the specific scenario developed by MAP focused on a potential upgrade path for the proton facilities at Fermilab, the fundamental accelerator concepts can readily be applied elsewhere.

## 2. Muon Production

Most muon facility designs developed in recent years are based on muon production as tertiary particles by decay of pions created with an intense, typically several MW, proton beam interacting in a heavy-material target. In order to achieve high luminosity in the collider, the resulting muon beam, produced with low energy and hence a limited lifetime, with very large transverse and longitudinal emittances, has to be cooled by approximately six orders of magnitude in the six-dimensional phase-space. It then has to be accelerated rapidly to mitigate muon decays. This is the standard scheme [6] that has been developed by MAP [1, 7] and is described in Section 2.1.

Recently, a novel approach based on muon pair production with a positron beam impinging on electrons at rest in a target [10] was proposed. This Low Emittance Muon Accelerator (LEMMA) study [11] is described in Section 2.2. Its principal advantage consists of providing beams with sufficiently small emittance that cooling of the beams is not required. This alternative concept exploits the fact that muons produced in $e^+e^-$ interactions close to threshold are constrained to occupy a small region in phase-space, thus producing a muon beam with a small emittance and long laboratory-lifetime due to the boost of the muons in the laboratory frame. A very small emittance can be obtained with this scheme [12], comparable to that typically achieved with electron beams. This opens up the possibility of obtaining high luminosity with relatively small muon fluxes, thus reducing background rates and activation problems due to high energy muon decays. For MC applications, very intense positron beams must be accumulated in a storage ring with a suitable internal target for muon production. The choice of the target is one of the crucial aspects for the success of this novel technique. Other key issues include the development of appropriate accelerator optics for the positron storage ring and the development of the required techniques for achieving and maintaining high muon rates.

### 2.1. *A Proton-Driven Muon Source*

#### 2.1.1. *Design Status*

In the MAP scheme [1, 7], muons are produced as tertiary particles by decay of pions created by a high-power proton beam impinging on a high-Z material target. The majority of the produced pions have momenta of a few hundred MeV/c, with a large momentum spread, and transverse momentum components that are comparable to their longitudinal momentum. Hence, the daughter muons are produced at low energy with a large longitudinal and transverse spread in phase-space. This initial muon population must be confined transversely, captured longitudinally, and have its phase-space manipulated to fit within the acceptance of an accelerator. These beam manipulations must be done quickly, before the muons decay with a lifetime at rest of $\tau_0 = 2.2$ μs.

Schematic layouts of muon-based NF and MC facilities are sketched in Fig. 2. The functional elements of muon beam generation for each proton-driven facility are very similar, thus providing a large number of facility synergies that can be leveraged for a productive scientific research program. For facilities based on protons, these elements are:

- A proton driver producing a high-power multi-GeV bunched H-beam. The primary requirement is the number of useful muons produced at the end of the decay channel, which, to good approximation, is proportional to the primary proton beam power, and (within the 5–15 GeV range) only weakly dependent on the proton beam energy. Considering a conversion efficiency of about 0.013 muons per proton-GeV [13] a proton beam in the 1-4 MW power range at an energy of 6.75 GeV provides the number of muons of each kind required for NF or MC applications.

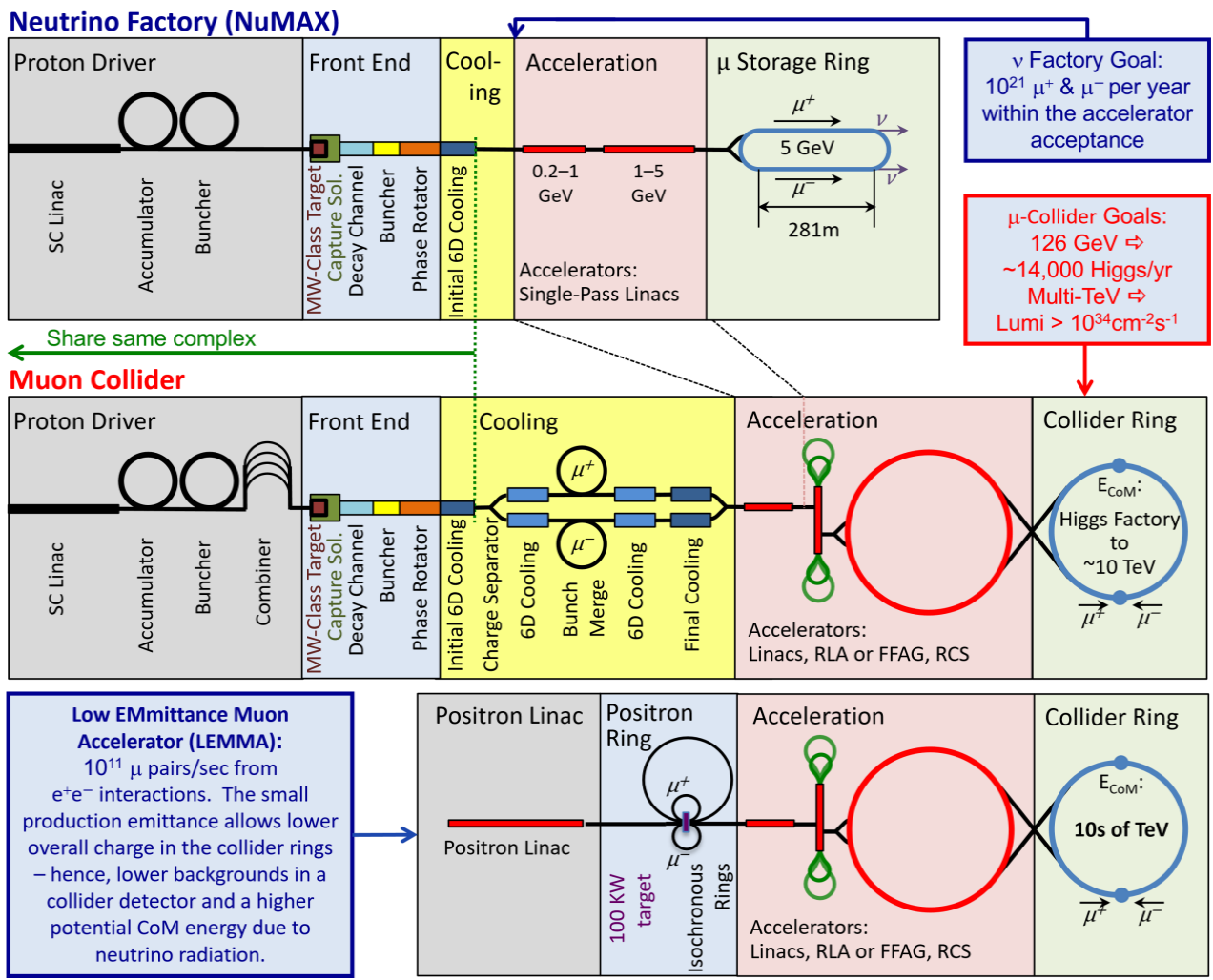

Figure 2. Schematic layout of NF and MC complexes based on the proton driver source scheme and on the low emittance positron source scheme.

- A buncher comprised of an accumulator and a compressor that forms intense and short (~2 ns) proton bunches. The first storage ring (the Accumulator) accumulates the protons via charge stripping of an H⁻ beam. The incoming beam from the linac is chopped to allow clean injection into pre-existing RF buckets. The second storage ring (the Compressor) accepts two to four bunches from the Accumulator and then performs a 90º bunch rotation in longitudinal phase space, shortening the bunches to the limit of the space-charge tune shift just before extraction. The Compressor ring must have a large momentum acceptance to allow for the beam momentum spread (a few %) during bunch rotation. For collider applications, working in single bunch colliding mode for luminosity optimization, the short bunches are extracted from the Compressor into separate ("trombone") transport lines of differing lengths so that they arrive on the pion production target simultaneously.
- A pion production target [14], capable of withstanding the high proton beam power, inserted in a high field solenoid to capture the pions and guide them into a decay channel.
- A front-end [13, 15] comprised of a solenoid decay channel, equipped with RF cavities, that captures the muons longitudinally into a bunch train and then applies a time-dependent acceleration that increases the energy of the slower (low-energy) bunches and decreases the energy of the faster (high-energy) bunches.
- An "initial" cooling channel that uses a moderate amount of ionization cooling [16, 17, 18, 19] to reduce the 6D phase space occupied by the beam by a factor 50 (a factor 5 in each transverse plane and a factor 2 in the longitudinal plane), so that it fits within the acceptance of the first acceleration stage.
- As presented in Section 3.1, further ionization cooling stages are necessary for high luminosity collider applications to reduce, by up to five orders of magnitude, the 6D phase space occupied by the beam from the initial volume at the exit of the front end to the parameters required by the collider. The evolution of the emittance, from initial production through the Final Cooling stage required for a high energy collider, is shown in Fig. 3.
- The beam is then accelerated with a series of fast acceleration stages, which may include Recirculating Linear Accelerators (RLA) [20], Fixed Field Alternating Gradient (FFAG) machines [21], and/or Rapid Cycling Synchrotrons (RCS) [22]. The accelerator chain will take the muon beams to the desired energy before injection in the NF Storage Ring or the MC Ring.

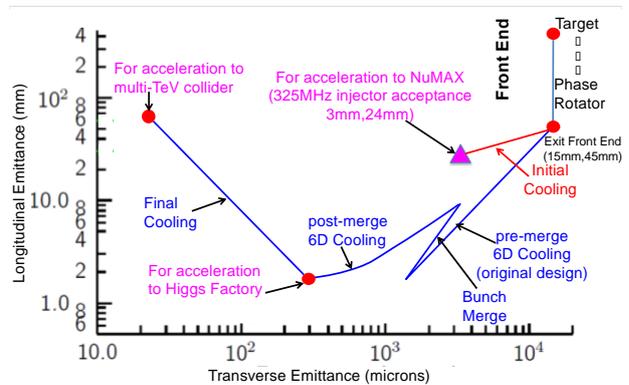

Figure 3. Ionization Cooling path in the 6D phase space.

### 2.1.2. *R&D Status*

The MAP R&D program focused on addressing those issues that would establish the feasibility of constructing and operating the systems required for a Neutrino Factory or Muon Collider to accomplish the production, capture, acceleration and storage of muon beams. A key part of the approach was to identify the minimum performance parameters that would enable key scientific thrusts to be executed. For instance, in the case of the proton-driver target, this led to the clear conclusion that a 1 MW target would be sufficient to provide the initial performance required by a long baseline NF (see the NuMAX discussion in the next section). Furthermore, 1 MW target systems are well within the performance envelope of target development efforts that are presently underway for facilities such as LBNF-DUNE. Given that the MERIT experiment [14] demonstrated the potential to handle several MW of incident proton power, it was concluded that further feasibility R&D in this area was not required (although significant engineering R&D would still be necessary to achieve a final target design at multi-MW power levels).

The above analysis led the MAP team to identify 5 areas where technology R&D to demonstrate feasibility of the accelerator concepts based on a proton-driver source was required. These 5 areas were:

- Operation of RF cavities in high magnetic fields for the buncher and phase rotator sections of the Front End as well as for the Cooling Channel;
- Providing a 6D Cooling lattice design that was consistent with the demonstrated performance parameters of the required magnets, absorbers and RF cavities;
- A measurement of ionization cooling in the correct momentum regime for a muon cooling channel – this was the goal of the International Muon Ionization Cooling Experiment (MICE);
- For a high energy MC, demonstration of high field solenoids and associated lattice suitable for Final Cooling;
- Demonstration of fast ramping magnets that would enable an RCS capability for accelerating muon beams to the TeV scale.

Since 2013, significant R&D progress was achieved so that each of these feasibility issues can be considered to be fully or nearly fully addressed.

*Operation of RF Cavities in High Magnetic Fields:*
A very successful program of research into the operation of RF cavities in high magnetic fields was completed at the Fermilab MuCool Test Area (MTA) in 2017. This program first demonstrated the operation of cavities filled with hydrogen gas that could achieve the accelerating gradients required for use in an ionization cooling channel while immersed in a 3 T magnetic field [23]. A second key demonstration was the successful testing of the final prototype RF module for the MICE experiment [24]. The module was intended for operation in vacuum and had undergone extensive RF design optimization for operation in magnetic fields. In tests at the MTA, it readily exceeded the operating specifications required by the MICE cooling channel design that included RF re-acceleration. Unfortunately, the final proposed step in the MICE program, which would have included operation of such modules, was cancelled. Finally, tests were completed in 2017 using an 805 MHz, in-vacuum cavity design with Beryllium end plates, which delivered accelerating gradients in excess of 50 MV/m in a 3 T magnetic field [25]. These three demonstrations span the design space required for the construction of an ionization cooling channel.

*Initial and 6D Cooling Lattice Designs:*
A core element of the MAP program was a design effort to achieve realizable cooling lattices, which fully incorporated the necessary engineering constraints associated with close integration of high-field solenoids, RF cavities and the necessary distribution of absorber material to enable the ionization cooling process. Design concepts were developed for Initial Cooling [26], 6D Cooling with RF cavities operating in vacuum (VCC) [17], a variant on this design where the cavities were assumed to be gas-filled [27] and the discrete absorber distribution optimized for this additional material in the channel (Hybrid) [28], and finally a Helical Cooling Channel (HCC) design that was optimized to operate as a gas-filled channel [29]. Each of these designs has utilized design parameters that are suitable for moving to a fully engineered technical demonstration of their capabilities. The overall simulated performance of the various designs relative to the MAP-established design targets are shown in Fig. 4. In the case of the vacuum RF cavity designs, the recent results from the MTA exceed the requirements for these lattices – thus opening up the possibility for further performance improvements for the VCC design. Additionally, the VCC and Hybrid designs, both of which assume the use of Low Temperature Superconductor (LTS) cables can potentially achieve even lower emittances by adding a final section utilizing High Temperature Superconductor (HTS) cable technology.

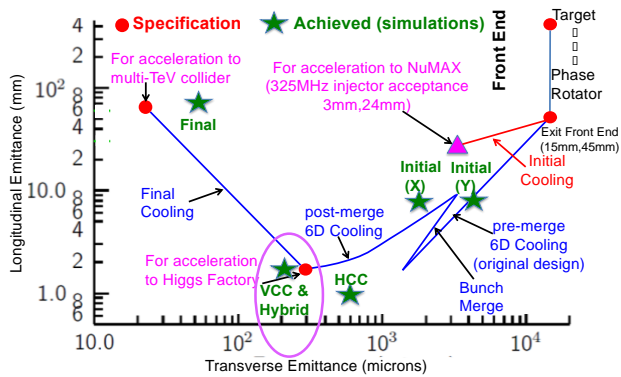

Figure 4. A version of Figure 3 with the results of the MAP design studies showing the performance achieved with full simulations of each of the cooling stages and methods. Given the recent technology results, solutions exist for the Initial and 6D Cooling stages that should meet the necessary performance requirements. While the Final Cooling channel misses its emittance target by just over a factor of 2, recent advances in high field solenoid construction have already exceeded the field and aperture requirements assumed in [30]**.**

*Measurements of Ionization Cooling:*

The International Muon Ionization Cooling Experiment (MICE) [31], hosted by Rutherford-Appleton Laboratory (RAL), undertook the task of characterizing the energy loss and multiple scattering characteristics of muons in the momentum regime relevant for the construction of an ionization cooling channel. These studies were carried out with the energy loss materials, LiH and $LH_2$, that offer the best performance for cooling channel designs. While descoping of the experimental program prevented a full cooling channel demonstration with RF re-acceleration, the experiment has successfully carried out detailed studies of the energy loss and multiple scattering characteristics of the above materials and has clearly measured the evolution of the emittance in the channel with results that are consistent with our models of cooling channel performance [32, 33]. The precise measurements of the muon interactions with the absorber materials will enable further refinement of our cooling channel simulations to precisely predict the expected performance of our designs. What is immediately clear is that the results validate the key assumptions and parameters being utilized in our present cooling channel designs.

*Very High Field Solenoid Magnet Development:*
The Final Cooling stage, as shown in Figs. 3 and 4, requires high field solenoids and a low energy linac to provide the emittance exchange that yields the small transverse emittances necessary for a high energy (TeV-class) collider design. We note that the MAP design study, which assumed a maximum solenoid field of 30 T with a 25 mm aperture, comes within roughly a factor of 2 of achieving the overall cooling performance goal [30,34]. A key driver in the performance of this section of cooling is the maximum magnetic field that can be assumed for the magnets. Thus, it should be noted that a recent demonstration of a 32 T, all superconducting user solenoid with 34 mm cold bore has been carried out at the National High Magnetic Magnet Laboratory (NHMFL) at Florida State University [35]. The rapid improvement in the demonstrated performance of HTS-based solenoids suggests that further optimization of the Final Cooling design is likely to deliver the required performance on an acceptable timescale for construction of the next high energy collider.

*Fast Acceleration to Collider Energies:*
A unique challenge for the MC is that the muons, with their short lifetime, must be accelerated very rapidly to high energies. To avoid excessive decay losses, the initial acceleration at low energies must be in linacs. Less costly options are required for high energy acceleration. In the MAP production scheme, where proton pulses arrive at a maximum rate of 15 Hz, pulsed rapid cycling synchrotrons (RCS) were identified as the most cost-effective solution with efficient power consumption [36]. In such a scheme, it was determined that fast ramping magnets capable of ±2 T peak-to-peak operation at a minimum of 400 Hz frequency were necessary to satisfy the lattice requirements. In the MAP designs, a hybrid RCS concept, where fixed field superconducting dipoles are interleaved with the fast ramping magnets, was employed. Initial studies with available magnet materials indicated that magnets meeting these requirements are achievable, although further work is required to prepare a full-scale working prototype [37].

## 2.2. *A Positron-Driven Muon Source*

The layout of a facility based on positron-driven muon beam generation (LEMMA) is compared in Fig. 2 to the one based on proton-driven muon beam generation (MAP). The most important properties of the muons produced by the positrons on target are: the low and tunable muon momentum in the CoM frame and the large boost of γ~200. These characteristics provide the following advantages: the final state muons are highly collimated and have very small emittance so that cooling of these beams is not required. These muons are produced with an average energy of 22 GeV corresponding to an average laboratory lifetime of ~500 μs, which also eases the acceleration scheme. The very small emittance of muons at production enables high luminosity with smaller muon fluxes reducing both the machine backgrounds in the experiments and more importantly the activation risks due to neutrino interactions.

The cross section for continuum muon pair production e$^+$e$^-$ → μ$^+$μ$^-$ has a maximum value of about 1 μb at $\sqrt{s}$ ~ 0.230 GeV. In the LEMMA scheme, the operational $\sqrt{s}$ is around 0.214 GeV. It is obtained from fixed target interactions with a positron beam energy of E$_+$ ~ s / (2m$_e$) ~ 45 GeV where m$_e$ is the electron mass.

The maximum scattering angle of the produced muons, θ$_\mu$(max), depends on the positron beam energy and, in the approximation of the muon velocity β$_\mu$=1, is given by

$$\theta_\mu(max) = \frac{2}{E_+}\sqrt{\frac{m_e E_+}{2} - m_\mu^2}. \quad (2)$$

The muons are produced with very small momentum in the rest frame and are contained in a cone of about 8×10$^{-4}$ rad for E$_+$ ~ 45 GeV. The energy distribution of the muons has an RMS distribution that increases with $\sqrt{s}$ from about 1 GeV at $\sqrt{s}$ =0.212 GeV to 2 GeV at $\sqrt{s}$ =0.214 GeV (E$_+$ ~ 45 GeV).

Muons are produced by positron beam annihilation with electrons at rest in the target. This target is the key component of the LEMMA scheme. It determines the muon beam quality as well as the production rate.

The number of μ$^+$μ$^-$ pairs produced per positron bunch interacting with a given target can be expressed as

$$n(\mu^+\mu^-) = N_b(e^+)\rho(e^-) L \sigma(\mu^+\mu^-) \quad (3)$$

where $N_b(e^+)$ is the number of positrons per bunch, $\rho(e^-)$ is the electron density in the medium, *L* is the target thickness, and $\sigma(\mu^+\mu^-)$ is the muon pair production cross section. We define the muon conversion efficiency as the ratio of the number of produced muon pairs to the number of incoming positrons:

$$\textit{eff}(\mu^+\mu^-) = \frac{n(\mu^+\mu^-)}{N_b(e^+)}. \quad (4)$$

The upper limit of this parameter is 10$^{-5}$, which is obtained for an ideal electron target where positron beam depletion is dominated by the radiative Bhabha process (positrons experiencing an energy loss of more than 2.5% will be below the dimuon production threshold).

On the one hand, it is desirable for a real target to be thin in order to minimize the emittance ($\varepsilon_{prod} \propto \theta_\mu(max)^2 \cdot L$). On the other hand, compact materials typically have a short radiation length. The short radiation length leads to an increase of the beam emittance due to multiple scattering as well as a rapid depletion of the positron beam due to bremsstrahlung.

Geant4 simulations show that the optimal targets must be thin and have low Z, such as carbon, beryllium, liquid lithium, or hydrogen. A dedicated study to identify the most suitable material is in progress. The best compromise to meet our requirements appears to be a 3 mm Be target, which provides

$$\textit{eff}(\mu^+\mu^-) = 7 \times 10^{-6}. \quad (5)$$

The very low muon production efficiency, due to the small production cross section, suggests a production scheme employing a positron ring with an internal target. This configuration will allow multiple interactions of the positron beam with the target for muon production. Fig. 5 shows the schematic layout of the Low Emittance Muon Accelerator (LEMMA): a 6 km positron ring with 100 bunches provides a rate of $1.5\times10^{18}$ positrons on target (T) per second, corresponding to $3\times10^{11}$ positrons per bunch. Muons are recombined in two $\mu^+\mu^-$ accumulator rings (AR) with a circumference of 60 m that corresponds to the positron ring bunch spacing. These rings intercept the positron ring at the interaction point with the target. Fig. 5 also shows the positron source with its adiabatic matching device (AMD).

2.2.1. *Design Status*

The LEMMA concept was first presented at Snowmass 2013 [10]. Initial studies [11] determined the main features of the muon beams and the layout scheme with a target in a 45 GeV low emittance positron ring.

The LEMMA positron ring was designed [12] with an emittance as low as $5.7\times10^{-9}$ m and a momentum acceptance of about 8%. The optical cell is based on the Hybrid Multi-Bend Achromat (HMBA) [38] to minimize emittance and maintain large momentum and dynamic acceptance. Table 1 summarizes the main parameters in the case of 32 regular cells with no IR for the target.

Table 1. Parameter table of the 45 GeV LEMMA positron ring

| Parameter | Unit | LEMMA e+ ring |
|---|---|---|
| Beam energy | GeV | 45 |
| Circumference | km | 6.301 |
| Geometric emittance x,y | m-rad | $5.7\times10^{-9}$ |
| No. e$^+$/bunch | $10^{11}$ | 3.15 |
| Number of bunches |  | 100 |
| Bunch length | mm | 3 |
| Trans. damping time | turns | 175 |
| Long. damping time | turns | 87.5 |
| Energy loss/turn | GeV | 0.511 |
| RF acceptance | % | +/- 7.2 |
| SR power | MW | 120 |
| RF frequency | MHz | 500 |
| RF Voltage | GV | 1.15 |

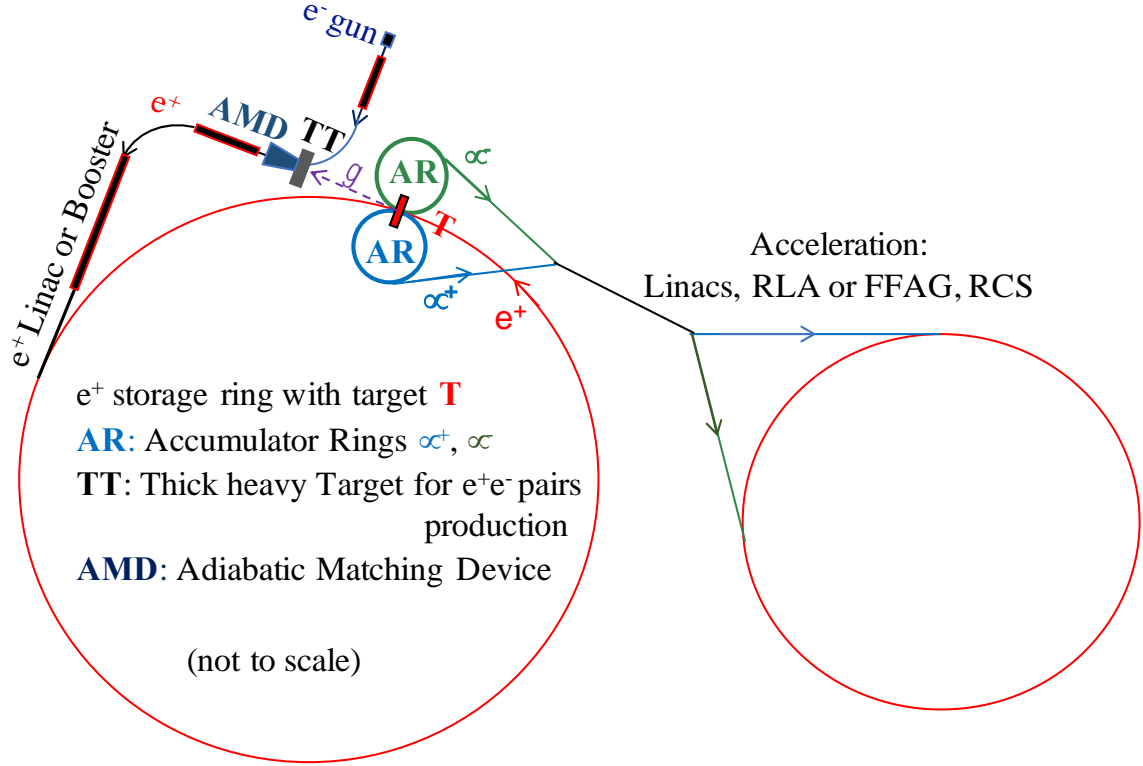

Figure 5. Positron-driven muon beam generation scheme.

Minimization of emittance growth in the positron beam is accomplished by controlling bremsstrahlung and multiple scattering at the target. This requires placing the target in a low-β and dispersion free location, similar to the interaction point in a collider. In addition, linear and non-linear terms related to momentum deviations must be minimized at the target to suppress emittance growth. This design has been validated by particle tracking simulations, where a β*=0.5 m with zero dispersion has been considered with a 3 mm Be target, finding a lifetime of about 40 turns [39]. A smaller value of β* has been explored to improve the emittance matching provided that the corresponding thermo-mechanical stress is found to be acceptable.

Numerical simulations are also explained analytically. At a beam waist after N machine turns the beam spot size is

$$\sigma_{x,y}(N) = \beta_{x,y}\sigma_{x',y'}(N), \quad (6)$$

where the beam divergence after N turns is given by

$$\sigma_{x',y'}(N) = \sqrt{\sigma^2_{x',y'}(0) + \sigma^2_{x',y'}(MS)}. \quad (7)$$

In this expression, the first term is the unperturbed beam divergence and the second one is the multiple scattering contribution after N turns

$$\sigma_{x',y'}(ms) = \sqrt{N}\theta_{r.m.s.}. \quad (8)$$

The final element of the study is to maximize the muon rate while preserving the best possible muon beam quality. The muon beam emittance is given by the convolution of multiple contributions:

$$\varepsilon(\mu) = \varepsilon_{prod} \oplus \varepsilon_{e^+} \oplus \varepsilon_{MS} \oplus \varepsilon_{Brems} \oplus \varepsilon_{AR}, \quad (9)$$

where $\varepsilon_{prod}$ is the muon production contribution, $\varepsilon_{e^+}$ is the positron beam contribution, $\varepsilon_{MS}$ is the multiple scattering contribution, $\varepsilon_{Brems}$ is the bremsstrahlung contribution, and $\varepsilon_{AR}$ is the contribution from the accumulator rings. Each of these values needs to be matched in each of the three phase space dimensions to minimize emittance growth due to beam filamentation. This can be achieved by matching the beam spots and beam divergence contributions at the target – i.e., each $\sigma_i$ and $\sigma_i'$ and their correlations in transverse phase space must be similar.

Constraints for the IR of the accumulator rings (AR) have been addressed [40] showing the need to match all contributions. The contribution of multiple scattering due to the multiple passages of muons through the target ($\varepsilon_{AR}$) is given by

$$\Delta\varepsilon_{AR} = \sigma_\theta^2 L/\sqrt{12}, \qquad (10)$$

where $\boldsymbol{\sigma_\theta}$ is the single passage scattering angle and L is the target thickness, where a perfectly matched beam phase space is assumed at the exit of the target.

For a 3 mm Be target the muon beam divergence after a single passage is increased by $\sigma_\theta = 59\ \mu rad$. Fig. 6 shows a simulation of the muon beam divergence as a function of turn in the AR for various positron beam energies and with a 3 mm Be target. After 2500 turns for a 45 GeV positron beam, $\sigma_\theta$ is about 1.8 mrad with a corresponding emittance increase $\Delta\varepsilon_{AR}$ of about 0.6 μm. The multiple scattering effect could be drastically reduced by using a crystal target in this challenging regime. In addition, studies to optimize the target performance with more conventional target materials, such as hydrogen and liquid lithium, are in progress.

The next major design effort for LEMMA will focus on the AR lattice design. The required parameter sets to support collider operation at CoM energies of 1.5, 3 and 6 TeV have been developed and a simulation campaign to optimize the overall emittance performance of the muon source will be carried out. Key R&D challenges are discussed in Section 4.1.2.

A scheme with a positron ring followed by a muon production multi-target system for an equivalent length of about 1 $X_0$ is also being considered, in analogy with the ERL scheme described in [11]. The main advantage is to increase the muon bunch population and possibly to reduce the thermo-mechanical stress on the target. Emittance preservation of positron and muon beams, the bunch recombination scheme as well as evolution of the parameter table is just being addressed and is presently under study.

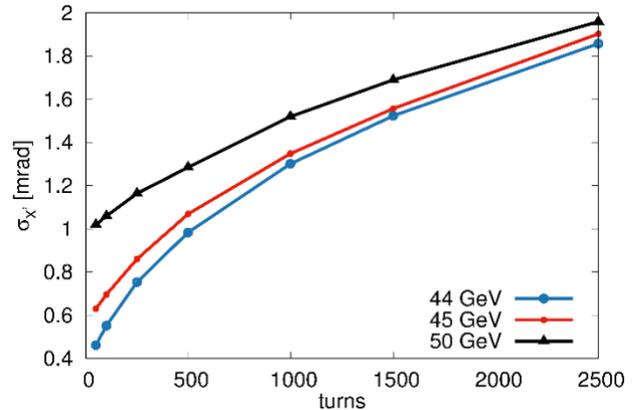

Figure 6. Beam divergence in the AR due to multiple scattering with a 3 mm Be target as a function of turn for 44, 45 and 50 GeV e+ beam energies.

## 3. From Source to Applications

Muon-based facilities [1, 7, 8] offer unique potential to provide next generation capabilities and world-leading experimental reach spanning physics at both the Intensity and Energy Frontiers. Muon accelerators can provide the next step with a high-flux and precise source of neutrinos to support a world-leading research program in neutrino physics. Furthermore, the infrastructure developed to support such an Intensity Frontier research program can also enable a subsequent stage of the facility that would support one or more stages of Muon Colliders. The MC could operate at CoM energies from the Higgs resonance at 126 GeV up to the multi-TeV scale. Their implementation in a laboratory already equipped with high power proton source or large positron flux would allow considerable savings. Sec. 3.1 describes a proton driver scenario optimized for the FNAL site taking advantage of existing or planned at the time equipment. It could be adapted to other laboratory sites. In particular, the US SNS facility with an available proton beam of up to 1.3-MW or the Japanese J-PARC facility recently upgraded up to 1 MW of proton beam power could be considered. The ESS facility in Sweden could provide an ideal proton beam with a power of up to 5 MW in the future and in parallel with the neutron program. Section 3.2 emphasizes the possible improvement opportunities provided by a positron driver scheme. Section 3.3 outlines the recent exploratory study of a Muon Collider in the CERN LHC tunnel.

### 3.1. *The MAP Staged Muon Facility Model*

In the MAP approach, an ensemble of facilities built in stages is made possible by the strong synergies between Neutrino Factories and Muon Colliders, both of which require a high-power proton source and target for muon generation followed by similar front-end and ionization cooling channels as emphasized in Fig. 2. These muon facilities rely on a number of systems with conventional technologies whose required operating parameters exceed the present state of the art as well as novel technologies unique to the MC. An R&D program to evaluate the feasibility of these technologies has been actively pursued within the framework of the U.S. Muon Accelerator Program (MAP) [7] and has already achieved impressive R&D results as summarized in Section 2.1.2.

#### 3.1.1. *Rationale for a Staged Approach*

The feasibility of the technologies required for Neutrino Factories and/or Muon Colliders must be validated before a facility based upon these capabilities can be proposed. Such validation is usually made in dedicated test facilities, which are rather expensive to build and to operate over several years. They are therefore difficult to justify and fund – this is particularly the case given that they are usually useful only for technology development rather than for physics research.

An alternative approach is considered here. It consists of a series of facilities, built in stages where each stage offers:

- Unique physics reach such that the facility supports a specific scientific need and can be funded based on its scientific output;
- An integrated R&D platform, in parallel with the physics program, that enables technology development, beam tests, and operational experience that can validate the technical solutions necessary for subsequent stages;
- Construction of each stage as an addition to the previous stages to ensure that prior investments are effectively utilized and such that the cost of each incremental stage remains acceptable.

#### 3.1.2. *The MASS Staging Scenario*

A complete staging scenario has been identified within the framework of the Muon Accelerator Staging Study (MASS) [2]. It consists of a series of facilities, each with performance characteristics providing unique physics reach as outlined in Fig. 7, which summarizes the physics reach of each element of the staging scenario. The parameters for the NF and MC applications are summarized in Table 2 and Table 3, respectively. The staged scenario is based on a progressive implementation of facilities with increasing complexity by adding systems to those previously installed. It leverages the strong synergies between NF and MC applications. Some stages could be skipped depending on the scientific thrusts being pursued. The proposed stages are:

- **nuSTORM** (**Neu**trinos from **STOR**ed **M**uons) [41]: a short-baseline NF-like ring enabling a definitive search for sterile neutrinos, as well as neutrino cross-section measurements that will ultimately be required for precision measurements at any long-baseline experiment.
- **NuMAX** (**N**eutrinos from a **M**uon **A**ccelerator Comple**X**) [42]**:** a long-baseline NF, initially operating with 5 GeV muons (optimized for CP violation measurements), but with sufficient flexibility to accommodate other energies as required by the science goals. The NuMAX ring would operate in conjunction with a far detector at a distance of ~1300 km, as envisaged at FNAL for LBNF/DUNE using SURF to house a deep underground detector:
  o An **initial (commissioning) phase** based on a limited proton beam power of 1 MW on the

muon production target. This phase would not employ 6D muon ionization cooling in order to provide an early and realistic startup configuration requiring only conventional technology, while still providing attractive physics parameters.

o The **NuMAX baseline**, which is upgraded from the commissioning phase by adding a limited amount of 6D cooling, thus providing a precise and well-characterized neutrino source that exceeds the capabilities of conventional superbeams.

o **NuMAX+**: a full-intensity NF, upgraded progressively from NuMAX by increasing the proton beam power on target as it becomes available, and simultaneously upgrading the detector for performance similar to the IDS-NF [43], as the ultimate source to enable precision CP-violation measurements in the neutrino sector and to characterize potential BSM physics discoveries.

- **Higgs Factory** [44, 45, 46]: a collider capable of providing between 3500 (startup) and 13,500 Higgs events per Snowmass year ($10^7$ sec) with exquisite energy resolution enabling direct Higgs mass and width measurements.
  o A further upgrade to the top quark production threshold could provide a **Top Factory** with production of up to 60000 top particles per Snowmass year ($10^7$ sec) for precise measurement of the top quark properties.
  o Further advances in 6D cooling performance could enable significantly improved luminosity performance for both the Higgs and Top Factory.
- **Multi-TeV Collider** [46, 47, 48]: if warranted by LHC results, a multi-TeV MC, with an ultimate energy reach up to roughly 10 TeV, likely offers the best performance and least cost and power consumption of any lepton collider operating in the multi-TeV regime.

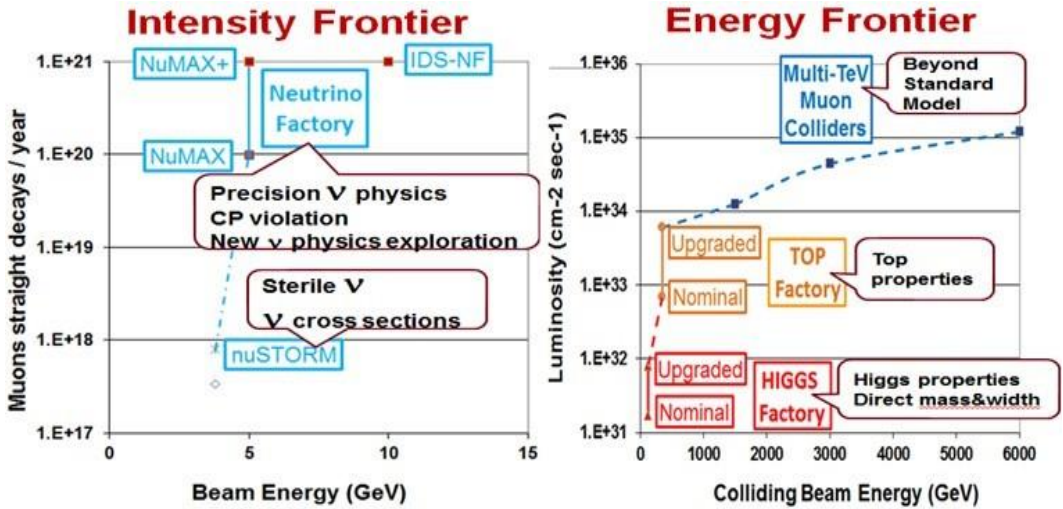

Figure 7. Performance and physics of muon-based facilities over two frontiers and a wide energy range.

Table 2. Main Parameters of the nuSTORM short-baseline NF and the three phases envisioned for the NuMAX long-baseline NF.

| System | Parameters | Unit | nuSTORM | NuMAX Commissioning | NuMAX | NuMAX+ |
|---|---|---|---|---|---|---|
| Performance | Stored $\mu^+$ or $\mu^-$/year | | $8\times10^{17}$ | $1.25\times10^{20}$ | $4.65\times10^{20}$ | $1.3\times10^{21}$ |
| | $\nu_e$ or $\nu_\mu$ to detectors/yr | | $3\times10^{17}$ | $4.9\times10^{19}$ | $1.8\times10^{20}$ | $5.0\times10^{20}$ |
| Detectors | **Far Detector** | Type | SuperBIND | MIND/Mag LAr | MIND/Mag Lar | MIND/Mag LAr |
| | Distance from Ring | km | 1.9 | 1300 | 1300 | 1300 |
| | Mass | kT | 1.3 | 100/30 | 100/30 | 100/30 |
| | Magnetic Field | T | 2 | 0.5-2 | 0.5-2 | 0.5-2 |
| | **Near Detector** | Type | SuperBIND | Suite | Suite | Suite |
| | Distance from Ring | m | 50 | 100 | 100 | 100 |
| | Mass | kT | 0.1 | 1 | 1 | 2.7 |
| | Magnetic Field | T | Yes | Yes | Yes | Yes |
| Neutrino Ring | Ring Momentum ($P_\mu$) | GeV/c | 3.8 | 5 | 5 | 5 |
| | Circumference (C) | m | 480 | 737 | 737 | 737 |
| | Straight Section | m | 184 | 281 | 281 | 281 |
| | Number of Bunches | - | - | 60 | 60 | 60 |
| | Charge per Bunch | $1\times10^9$ | - | 6.9 | 26 | 35 |
| Acceleration | Initial Momentum | GeV/c | - | 0.25 | 0.25 | 0.25 |
| | Single-pass Linacs | GeV/c | - | 1.0, 3.75 | 1.0, 3.75 | 1.0, 3.75 |
| | SRF Frequencies | MHz | - | 325, 650 | 325, 650 | 325, 650 |
| | Repetition Frequency | Hz | - | 30 | 30 | 60 |
| Cooling | Horizontal/Vertical/Longitudinal | | None | None | 5/5/2 | 5/5/2 |
| Proton Source | Proton Beam Power | MW | 0.2 | 1 | 1 | 2.75 |
| | Proton Beam Energy | GeV | 120 | 6.75 | 6.75 | 6.75 |
| | protons/year | $1\times10^{21}$ | 0.1 | 9.2 | 9.2 | 25.4 |
| | Repetition Rate | Hz | 0.75 | 15 | 15 | 15 |

Table 3. Main Parameters of the various phases of a Muon Collider as developed by the MAP effort.

| Parameter | Units | Higgs | *Top - High Resolution* | *Top - High Luminosity* | *Multi-TeV* | | |
|---|---|---|---|---|---|---|---|
| CoM Energy | TeV | 0.126 | 0.35 | 0.35 | 1.5 | 3.0 | 6.0* |
| Avg. Luminosity | $10^{34}cm^{-2}s^{-1}$ | 0.008 | 0.07 | 0.6 | 1.25 | 4.4 | 12 |
| Beam Energy Spread | % | 0.004 | 0.01 | 0.1 | 0.1 | 0.1 | 0.1 |
| Higgs Production/$10^7$sec | | 13,500 | 7,000 | 60,000 | 37,500 | 200,000 | 820,000 |
| Circumference | km | 0.3 | 0.7 | 0.7 | 2.5 | 4.5 | 6 |
| Ring Depth [1] | m | 135 | 135 | 135 | 135 | 135 | 540 |
| No. of IPs | | 1 | 1 | 1 | 2 | 2 | 2 |
| Repetition Rate | Hz | 15 | 15 | 15 | 15 | 12 | 6 |
| $\beta^*_{x,y}$ | cm | 1.7 | 1.5 | 0.5 | 1 (0.5-2) | 0.5 (0.3-3) | 0.25 |
| No. muons/bunch | $10^{12}$ | 4 | 4 | 3 | 2 | 2 | 2 |
| Norm. Trans. Emittance, $\varepsilon_T$ | $\pi$ mm-rad | 0.2 | 0.2 | 0.05 | 0.025 | 0.025 | 0.025 |
| Norm. Long. Emittance, $\varepsilon_L$ | $\pi$ mm-rad | 1.5 | 1.5 | 10 | 70 | 70 | 70 |
| Bunch Length, $\sigma_s$ | cm | 6.3 | 0.9 | 0.5 | 1 | 0.5 | 0.2 |
| Proton Driver Power | MW | 4 | 4 | 4 | 4 | 4 | 1.6 |
| Wall Plug Power | MW | 200 | 203 | 203 | 216 | 230 | 270 |

*Accounts for off-site neutrino radiation

3.1.3. *nuSTORM*

The nuSTORM proposal [41] is an ideal entry level short-baseline NF, which does not require any new technology development before construction (Fig. 8). It is based on a very reasonable proton beam power on target of 200 kW and a ring with large momentum acceptance where muons from pion decay are captured in the first straight [49]. This design provides up to $1.7 \times 10^{18}$ $\mu^+$ stored in the muon decay ring per operational year and can generate a very attractive number of neutrino oscillation events in all channels. This facility would also enable a measurement of neutrino cross sections with high precision at the percent level.

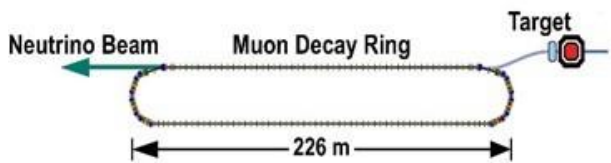

Figure 8.   nuSTORM layout

3.1.4. *NuMAX*

Preliminary parameters of the three NuMAX phases [42] with progressively increasing complexity and performance are presented in Table 2 and compared with the nuSTORM parameters. In particular, the final phase of NuMAX+ provides a neutrino flux similar to that obtained by the IDS-NF [43]. Its performance [35] compares favorably with those of other facilities as shown in Figure 9.

The NuMAX muon production system corresponds to the one described in Section 1.2. It is based on a proton driver accelerating an H⁻ beam up to 3 GeV, accumulated, compressed and further accelerated as proton bunches by a 650 MHz dual linac before hitting the target for pion production. The muons are produced by pion decay, captured and bunched in the front end, and are recirculated to the dual-use linac for further acceleration up to 5 GeV as required by NuMAX and sketched in Fig. 10.

The dual-use linac concept, which accelerates both the proton and muon beams, provides an opportunity for considerable savings. It requires initial ionization cooling [16] to match the muon beam emittances to the linac acceptances at the 325 and 650 MHz RF standards. The initial cooling specifications result from a cost optimization as the best trade-off between linac, RF and cooling. A fair comparison with a more conventional acceleration scheme based on separate linacs optimized to each species would have to be made to check if the possible cost savings of a dual-use linac are worth its additional challenges.

In order to achieve the required flux of $5 \times 10^{20}$ neutrinos per year at the far detector, 60 bunches of $3.5 \times 10^{10}$ muons/bunch are stored in the muon decay ring at a 15 Hz repetition rate. Assuming a reasonable overall muon transmission efficiency, including decay losses, along the NuMAX complex and a production of 0.08 useful muons per 6.75 GeV proton on target [13], the facility requires a high but not unreasonable proton beam power of 2.75MW on target for muon production. A modest amount of 6D cooling by a factor of 50 (5 in each transverse plane and 2 in the longitudinal direction) allows matching of the muon beam emittances to the acceptances of the cost-effective and power-efficient acceleration system.

The early NuMAX commissioning phase, which assumes no muon cooling and only 1 MW of proton beam power, corresponding to the present state of the art, already provides an attractive flux that is only one order of magnitude lower than that provided by the IDS-NF design. The flux is then improved by roughly a factor 4 by implementing the 6D cooling in the NuMAX baseline phase, while still maintaining 1 MW of proton beam power on the target. Finally, the flux can be further improved in the NuMAX+ configuration by increasing the proton power on target to 2.75 MW.

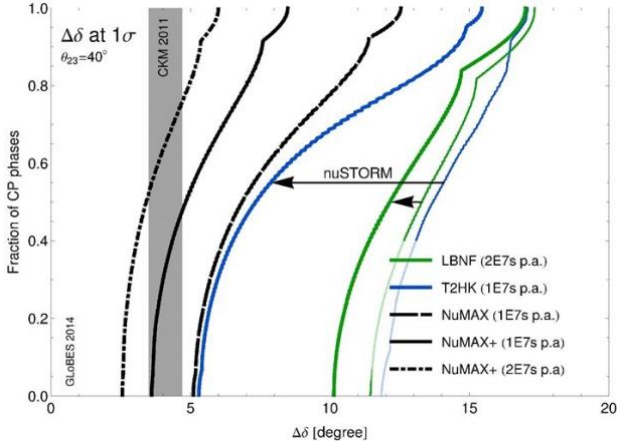

Figure 9. Physics performance [3] of the NuMAX stages in terms of CP-violating phase δ compared with the anticipated performance of LBNF and T2HK. The benefits to LBNF and T2HK of precision neutrino cross section measurements that can be obtained with nuSTORM [41] are also shown.

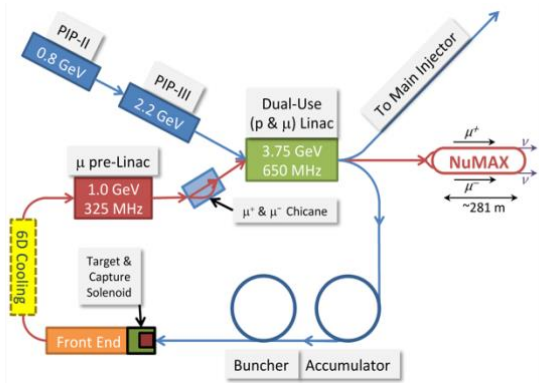

Figure 10. Layout of a muon-based NF as envisioned by MAP, which was based on the proton upgrade path described by the Project X proposal for Fermilab [50].

### 3.1.5. *Higgs/Top Muon Collider*

A Higgs MC [44] as sketched in Fig. 11 could naturally follow a NF complex by taking advantage of all systems already installed for the NF and adding the specific equipment necessary to support the collider. In particular, the muon generation system is very similar to that of a NF with the following limited exceptions:

- a combiner to simultaneously deliver multiple proton beams on target in order to provide the necessary production rates to deliver the desired collider luminosity;
- the proton driver and target upgraded to utilize a 4 MW proton beam;
- an additional ionization cooling stage to reduce the beam emittance to the level required by the collider for luminosity performance.

In order to take advantage of the large cross section for Higgs production at the s-channel resonance, the colliding beams in a Higgs factory require a very small energy spread and excellent energy stability (a few parts in $10^5$). The ionization cooling system of the Higgs Factory is therefore chosen to minimize the beam momentum spread on the cooling path shown in Fig. 3, corresponding to normalized emittances of 200 μm in the transverse and 1.5 mm in the longitudinal planes. It therefore requires an additional high performance 6D ionization cooling channel, beyond that of the NF, to reduce the emittance of both $\mu^+$ and $\mu^-$ beams by a factor of 3600 (15 in each transverse plane and 16 in the longitudinal plane).

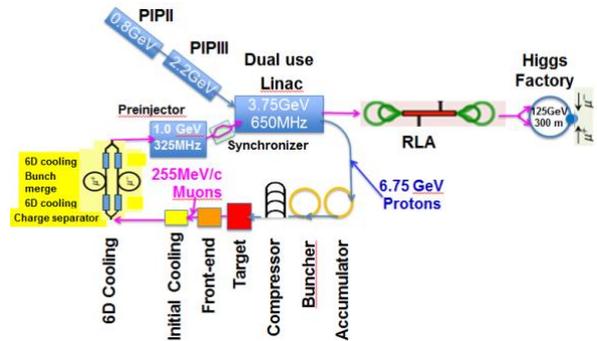

Figure 11. Layout of a muon-based Higgs Factory.

The 5 GeV $\mu^+$ and $\mu^-$ beams at the exit of the dual-use linac are further accelerated up to the colliding beam energy of 62.5 GeV by a series of fast accelerators in order to minimize muon decays during acceleration. Our baseline assumption is the use of an RLA [51], although FFAG [21] or RCS [22] solutions can also be considered.

The layout and optics of a Higgs collider ring [45] with a single interaction point and detector is shown in Fig. 12. By equipping the ring with 10 T superconducting dipole magnets, the ring circumference is limited to ~300 m. This allows the muons to circulate in the ring for >1000 turns before decaying, thus maximizing the number of collisions and the integrated luminosity that can be achieved.

The detector of a MC is likely to share many key technologies with other modern colliders, including those designed for high energy measurements of $e^+e^-$ collisions at linear colliders or protons at the LHC. However, the background considerations are quite different due to the contribution of muon decays in the collider ring. Special shielding designs are required to help control these backgrounds in the detector [52].

The main ring and beam parameters for MC designs at a range of CoM energies from the Higgs pole to the multi-TeV scale are summarized in Table 3. In the Higgs Factory configuration, the MC provides an attractive number of Higgs particles typically 27,000 for an operational year of $2 \times 10^7$ s, where new advances in 6D cooling concepts could further increase this rate. We anticipate that the luminosity at startup would be about a factor of 4 lower than the baseline due to the reduced bunch charge that would initially be available.

When a clearly defined scientific case for a multi-TeV lepton collider is in hand, a low energy MC could be upgraded to a TeV-class collider with a layout as shown in Fig. 13. The multi-TeV facility would reuse all of the systems already installed and would add the specific systems required for TeV-scale operation. In particular a final cooling stage would be deployed that enables an exchange between the transverse and longitudinal beam emittances in order to provide the small transverse beam sizes required for the targeted luminosity performance. As shown on the ionization cooling path displayed in Fig. 3, the normalized transverse emittances of the muon beams are reduced in a final cooling stage by one order of magnitude, to 25 μm, at the expense of a longitudinal emittance increase to 70 mm.

The $\mu^+$ and $\mu^-$ beams are further accelerated to the required colliding beam energy by the addition of one or more rapid cycling synchrotron stages [22], which benefit from the increased muon lifetime in the laboratory frame as the beam energy increases.

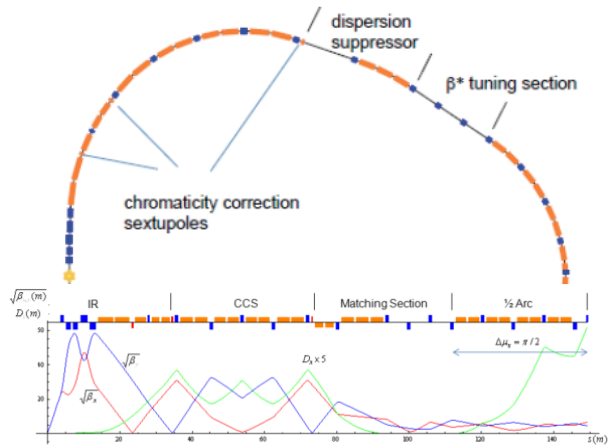

Figure 12.  Higgs Factory half-ring and lattice functions for a single interaction region with $\beta^*=2.5$ cm.

As noted previously, a Higgs Factory could be followed by a Top Factory operating at the top quark production threshold by increasing the beam energy to 175 GeV per beam with an optimized collider ring circumference of 700 m.

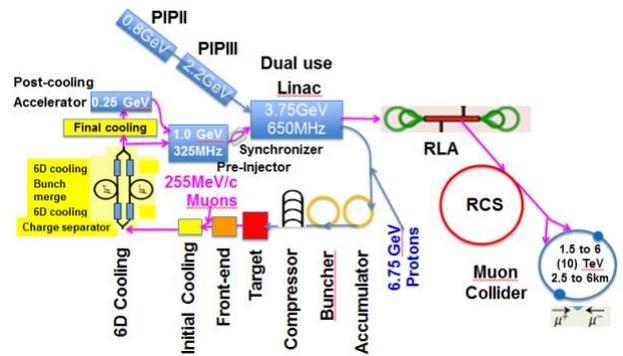

Figure 13.  Layout of a muon-based multi-TeV Muon Collider

The MC parameters for collider rings at 1.5 TeV, 3 TeV [47] and 6 TeV [48] are summarized in Table 3. Assuming that the main dipoles are 10 T superconducting magnets, the respective circumferences of 2.5, 4.5 and 6 km allow multiple collisions in two detectors for roughly 1000 turns during the lifetime of the muon beams.

The detector of a multi-TeV MC is extrapolated from that of the lower energy collider and must account for the higher energy of the muon beams. Although the energy of the decay products increases with the muon beam energy, the absolute rate that must be dealt with decreases due to the increased muon lifetime in the lab frame and the smaller ring curvature which mitigates the number of muon decays impacting the detector.

Site radiation issues potentially set a limit on the average beam current in a MC as the CoM energy increases [1]. In order not to exceed an off-site radiation level of 0.1 mSv (10 mR) per year, the collider ring must be located deep underground (a ring depth of 135 m, similar to the LHC, appears reasonable up to a CoM energy of ~3 TeV with greater depths required for higher energies). In addition, the average collider beam current in the MAP designs is limited by reducing the beam repetition rate to 12 Hz at 3 TeV and 6 Hz at 6 TeV, which also benefits the facility power consumption. As a consequence, the beam power required from the proton driver of a 6 TeV MC is reduced to 1.6 MW, which is very close to the present state of the art for high power linacs and targets.

### 3.1.6. *Wall-Plug Power Estimation*

The power consumption of the various accelerator and detector systems of a 1.5, 3 and 6 TeV MC has been estimated in [1]. For a fair comparison, after adding power for the conventional facilities similar to that estimated in other more mature projects, the power consumption of various lepton colliders as a function of the colliding beam energy is compared in Fig. 14. It can be seen that a MC requires significant base power of roughly 100 MW, independent of the colliding beam energy, for the muon production. The power consumption then increases with colliding beam energy due to the additional RF power required for beam acceleration and the high energy combiner ring. Nevertheless, the power increase with colliding beam energy is much lower than for $e^+e^-$ linear colliders because of the efficiency inherent in multi-pass acceleration in rings.

As displayed in Fig. 1, the figure of merit of muon-based facilities, defined as the luminosity produced per unit of wall-plug power, is increasing with the colliding beam energy. At colliding beam energies lower than roughly 2 TeV, the power consumption of $e^+e^-$ linear colliders and $\mu^+\mu^-$ colliders based on a proton driver source is comparable. The figure of merit of $e^+e^-$ circular colliders is larger than any other technology due to their efficient beam acceleration and multiple collisions in rings. But it rapidly decreases with the colliding beam energy, which is limited to about 350 GeV even in very large ring like LHC or FCC due to the large power emitted as synchrotron radiation. At colliding beam energies greater than 2 TeV, the MC figure of merit significantly exceeds that of any other lepton collider technology because of the efficient acceleration and multi-turn collisions in the rings. This is even more true for the luminosity of colliding particles with a small momentum spread since the luminosity of $e^+e^-$ collisions is limited by beamstrahlung which does not affect muon collisions. As a consequence, a MC appears to provide the most effective technology option for lepton colliders in the multi-TeV range above 2 TeV.

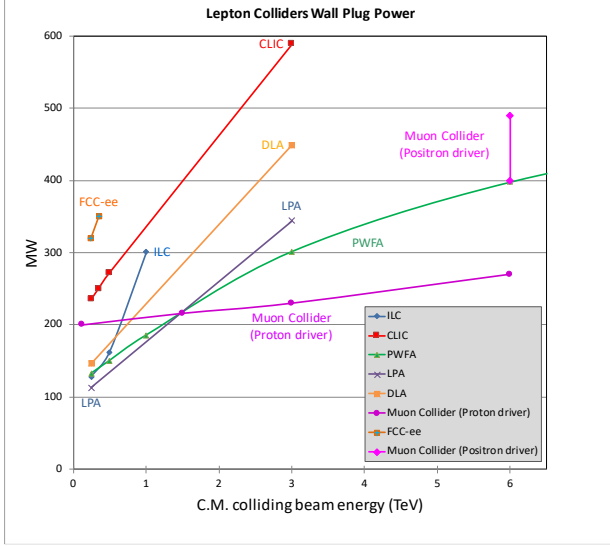

Figure 14. Power consumption of lepton colliders using various accelerator technologies.

## 3.2. *Opportunities with LEMMA*

### 3.2.1. *Potential for a Multi-TeV Collider*

A muon-based Higgs Factory requires a very small muon beam energy spread, on the order of a few parts in $10^5$. The LEMMA scheme provides a muon beam energy spread that depends on the positron beam energy $E_+$ in the positron storage ring (i.e., the $e^+e^-$ CoM energy) as

$$\Delta E_\mu = \sqrt{\frac{E_+}{2m_e}}\sqrt{\frac{m_e E_+}{2} - m_\mu^2}, \qquad (11)$$

where $m_e$ and $m_\mu$ are the electron and muon mass and $E_+$ is the positron beam energy.

Although a natural beam energy spread for the produced muons of 0.5 % can be obtained close to the production threshold of 43.8 GeV, the energy spread required for effective s-channel Higgs production cannot be achieved directly using this production method. Furthermore, it does not appear feasible to utilize emittance exchange or momentum cooling at such high production energies to reduce the energy spread to the required values. Thus, the energy spread and luminosity for the LEMMA scheme are insufficient for a Higgs Factory capable of carrying out the line shape measurement for $\mu^+\mu^- \to H$.

On the other hand, the LEMMA scheme is very attractive at multi-TeV energy scales, where high luminosities can be achieved with significantly lower bunch charges – hence, with much lower muon production rates relative to a proton-driven source, thus mitigating the neutrino radiation issue for the site. Consequently, the LEMMA scheme could extend the energy reach of a MC which is usually limited by neutrino radiation. A very high energy MC would be a discovery machine, with a direct reach for new physics comparable to that of proton colliders such as the LHC or FCC-hh, and with high indirect reach on new physics. Such a machine would be very competitive with linear colliders such as CLIC.

A preliminary set of parameters for a MC with 6 TeV CoM energy based on the LEMMA concept is shown in Table 4 [53]. This table should be viewed as summarizing the goals of the LEMMA design study as opposed to parameters based on thoroughly evaluated design concepts.

A luminosity of $5\times 10^{34}$ cm$^{-2}$s$^{-1}$ would be obtained with collisions of single bunches containing $6\times 10^9$ muons with transverse emittances of 40 nm. A muon beam emittance as low as 40 nm will only be possible if there is a negligible contribution from multiple scattering through the target. With the present 3 mm beryllium target, this contribution is about a factor 15 times larger. Further R&D is necessary to determine an optimized and realistic value. This effort will need to consider various target options with careful optics matching at the target.

A muon bunch charge of $4.5\times 10^7$ is provided by the AR, whereas a bunch charge of $6\times 10^9$ muons is considered in the collider, as shown in Table 4. This enhancement by a factor ~120 (the ratio of the laboratory lifetime of a muon in the collider versus at production) will require a bunch combiner scheme at the collider energy that does not have significant adverse impact on the beam emittance [54, 55]. In such a scheme, a muon bunch with almost constant intensity will continuously circulate in the collider. This combiner approach might be realized either in the longitudinal [54] or in the transverse plane [55]. Detailed optics and beam dynamics studies will be essential to determine whether such a scheme is feasible and can be implemented without significant emittance growth, thus enhancing the luminosity performance.

The beta functions at the IP are as small as 0.2 mm with a round beam, thus implying a nano-beam scheme in the final focus. The lattice for the collider has not been designed yet, so the final focus parameters will be confirmed as part of the optics design. This small beta value requires a bunch length as small as 0.1 mm, posing collective effects and wake-fields constraints that will have to be addressed as well.

Table 4. Preliminary Parameters for a LEMMA-based 6 TeV Muon Collider.

| Parameter | Unit | LEMMA-6 TeV |
|---|---|---|
| Beam energy | TeV | 3 |
| Luminosity | $cm^{-2}s^{-1}$ | $5.1 \times 10^{34}$ |
| Circumference | km | 6 |
| Bending field | T | 15 |
| No. muons/bunch | # | $6 \times 10^9$ |
| No. bunches | # | 1 |
| Beam current | mA | 0.048 |
| Normalized Emittance x,y | m-rad | $40 \times 10^{-9}$ |
| Emittance x,y | m-rad | $1.4 \times 10^{-12}$ |
| $\beta_{x,y}$ @IP | mm | 0.2 |
| $\sigma_{x,y}$ @IP | m | $1.7 \times 10^{-8}$ |
| $\sigma_{x',y'}$ @IP | rad | $8.4 \times 10^{-5}$ |
| Bunch length | mm | 0.1 |
| No. turns before decay | # | 3114 |
| µ lifetime | ms | 60 |

### 3.2.2. *Neutrino radiation*

High energy muon rings generate significant neutrino radiation that potentially sets the maximum acceptable energy for a MC. As the muons decay in a collider ring, the resulting neutrinos form a fan with vertical height of the order of $1/\gamma$ that eventually breaks the earth's surface. The first paper that identified the problem of the neutrino radiation hazard in the extreme conditions of a high energy MC was by B. J King [56]. The low charge per bunch in the LEMMA scheme can provide a significant reduction in the radiation dose from this source.

Fig. 15 compares neutrino radiation doses for muon bunches with $3 \times 10^{13}$ particles/s, representative of a proton driver source, and $10^{11}$ particles/s, representative of a positron source, in a ring at 100 m depth (higher dose in straight sections) as from [57] where 8 T dipole fields are considered in the arcs. In the straight sections the dose depends on its length and is assumed to be a factor ten higher. For a more comprehensive estimate, formulas reported in Ref. [1] can be used.

### 3.2.3. *Lower Backgrounds*

The principal backgrounds in a MC are due to muon decay. The fraction of electrons and positrons exceeding the physical aperture and producing backgrounds depends on IR details that have not yet been studied for the LEMMA scheme. Assuming the same IR configuration as was studied for the proton-driver-based machines, a lower bunch intensity by a factor of 300 translates directly into a background reduction by the same factor. The impact and feasibility of an IR design with lower β remains to be addressed.

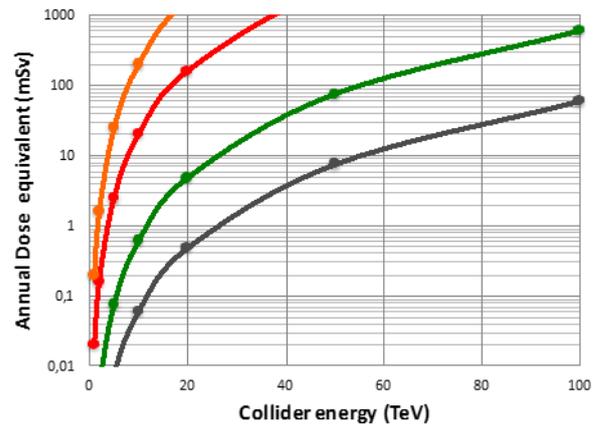

Figure 15. Dose equivalent due to neutrino radiation as a function of the collider energy (muon beam energy); for the MAP (orange and red) and the LEMMA (green and black) fluxes in the straight and bending sections, respectively [57].

3.2.4. *Power Requirements*

A portion of the 45 GeV positron beam is lost due to its interaction with the muon production target on each turn and must be replenished. This poses constraints on the specifications for the positron source and the wall plug power needed to accelerate the positrons. The wall plug power needed depends on specific aspects of the acceleration scheme. A first rough estimate can be obtained by noting that a 1% positron beam loss per turn, corresponding to about 100 MW that must be replaced. For this reason, the positron beam loss has to be minimized by increasing the positron ring momentum acceptance. The present positron ring lattice has positron losses of about of 3% per turn in simulation.

The synchrotron radiation power loss in the 6 km $e^+$ ring is about 120 MW. Studies are planned to increase the ring circumference to decrease the SR losses and to optimize the other parameters as described in Ref. [39]. In addition to this requirement on power for the LEMMA muon source, the power requirements of the AR, the fast acceleration systems and the MC ring need to be added to complete the wall plug power estimate for the LEMMA accelerator complex.

## 3.3. *A 14 TeV Muon Collider in the LHC Tunnel*

The technical feasibility of a pulsed 14 TeV CoM energy MC in the CERN LHC tunnel has recently been considered [58] as sketched in Fig. 16. It leverages the existing CERN facilities, including the 26.7 km circumference LHC tunnel and its injectors, in order to provide significant cost savings. Collisions of fundamental leptons at the specified energies could provide a physics reach comparable to the interaction of the proton constituents in a 100 TeV FCC-hh with lower cost and cleaner physics conditions.

The collider performance is determined by the intensity and brightness of the muon source. Table 5 summarizes key parameters of three options for a 14 TeV MC in the LHC tunnel with a beam-beam lifetime in collision of 146 ms: the first one adapts the existing 24 GeV CERN PS source, the second requires a new 8 GeV linac and storage ring [48], while the third is based on the threshold production of $\mu^+\mu^-$ as proposed in the LEMMA scheme.

The challenges already described for the MAP and LEMMA approaches still apply to this proposal. In particular, cooling performance consistent with the MAP design studies is required for the proton-based source options. The LEMMA-based design needs very serious optimization to ease the facility power requirements. Acceleration based on pulsed and CW SRF should generally be considered feasible for gradients of about 30 MV/m (pulsed) and 20MV/m (CW). The required pulsed magnets exist only in prototypes and considerable technology development is required to prove technical feasibility. The main attraction of such a 14 TeV $\mu^+\mu^-$ collider is its potential cost savings (see Fig. 4 of Ref. [58]). In this scheme, the civil construction costs can be reduced by reusing the existing 27 km LHC tunnel, the 7 km SPS tunnel and the accompanying CERN infrastructure. The MAP-based scenario provides a very attractive luminosity but at higher cost and at the limit of acceptable site radiation due to neutrinos. The PS-based scheme provides a luminosity two orders of magnitude lower, but with a lower cost and a much safer neutrino radiation level. It could be considered as a first stage scenario. The neutrino radiation of the LEMMA-based scheme is negligible but its luminosity is lower by another order of magnitude due to the limitations in the muon bunch intensity.

In contrast, the performance listed in Table 4 for a 6 TeV MC makes use of the full muon rate available from the LEMMA scheme, where it is assumed bunch combination can be achieved with no luminosity degradation [54, 55]. The implementation of a LEMC can be considered the final phase of a long-term prospective lepton accelerator facility at CERN starting from $e^+e^-$.

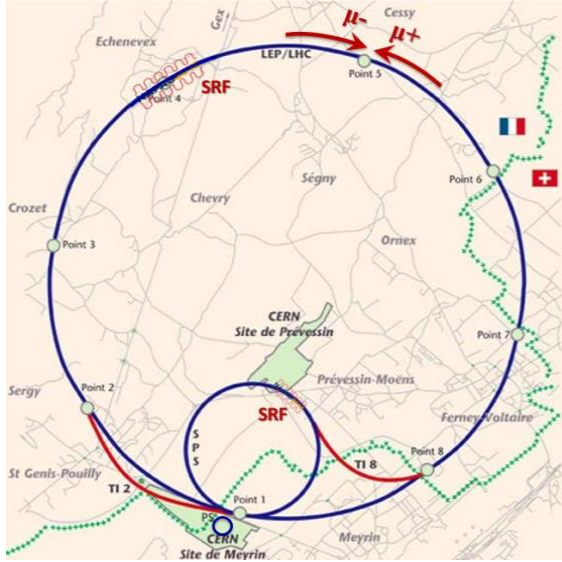

Figure 16. Schematic layout of a pulsed MC at a beam energy of 7 TeV in the LHC tunnel [58] with modifications indicated in red.

Table 5. Options for a 14 TeV $\mu^+\mu^-$ Collider in the LHC Tunnel from Ref. [58].

| Parameter | Units | "PS" | "MAP" | "LEMC" |
|---|---|---|---|---|
| Avg. Luminosity | $cm^{-2}s^{-1}$ | $1.2\times10^{33}$ | $3.3\times10^{35}$ | $2.4\times10^{32}$ |
| Energy Spread | % | 0.1 | 0.1 | 0.2 |
| Rep Rate, Hz | Hz | 5 | 5 | 2200 |
| $N_\mu$/bunch | # | $1.2\times10^{11}$ | $2\times10^{12}$ | $4.5\times10^7$ |
| $n_b$ | # | 1 | 1 | 1 |
| $\varepsilon_{LN}$ | mm-mrad | 25 | 25 | 0.04 |
| $\beta^*$ | mm | 1 | 1 | 0.2 |
| $\sigma^*$ | µm | 0.6 | 0.6 | 0.011 |
| Bunch Length | mm | 0.001 | 0.001 | 0.0002 |
| µ Production Source | | 24 GeV $p$ | 8 GeV $p$ | 45 GeV $e^+$ |
| $P$ or $e^+$/pulse | | $6\times10^{12}$ | $2\times10^{14}$ | $3\times10^{13}$ |
| Driver beam power | MW | 0.17 | 1.6 | 40 |
| Acceleration | GeV | 1-3.5 3.5-7 RCS | 1-3.5 3.5-7 RCS | 40GV, RLA 20 turn |
| ν Radiation | mSv/yr | 0.08 | 1.5 | 0.015 |

## 4. Perspectives on Future R&D

This section describes our view on priority areas that should be targeted for continued R&D and sub-system demonstrations in order to provide confidence in accelerator technologies based on muon beams. We hope that this provides useful input for the community as it lays out its plans to assess future collider and neutrino beam options.

### 4.1. *Source*

#### 4.1.1. *Proton Driver Target*

A key conclusion of the MAP effort was that muon accelerator capabilities for both a NF and MC could initially be supported by a target system operating at 1 to 2 MW of incident proton power. Furthermore, the baseline design assumptions for the highest energy collider assumed a reduced power on target from the highest values to ensure that site radiation issues could be adequately handled. With these elements of the overall plan, the proton driver power requirements versus the capability being provided are shown in Fig. 17.

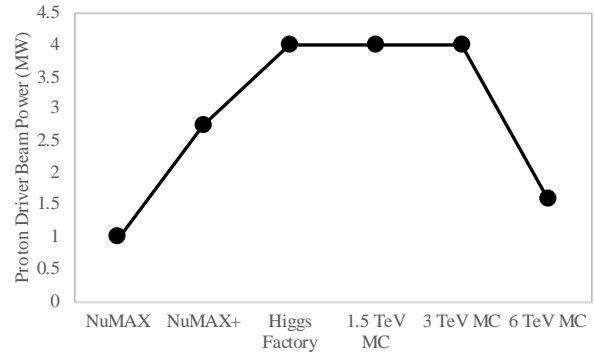

Figure 17. Baseline power requirements for muon accelerator capabilities based on a proton driver source.

Because of the engineering challenges associated with solenoid-based capture of the pions generated in the target, an effort to move forward with additional work on the capture section conceptual design would be highly valuable.

#### 4.1.2. *R&D for the Positron-driven Source*

*Muon Production Target*
A systematic research effort on the muon production target for the positron-driven source is required. In particular an optimized choice of the target material is needed. Initial beam dynamics simulations using Geant4 and G4-beamline suggest beryllium as the best solution. However, carbon composites, as well as other candidate materials (e.g., liquid lithium and liquid hydrogen), need to be studied. This effort must include an assessment of the material's mechanical stress and heat load resistance properties with proper engineering simulations. Experimental tests with candidate target materials will be required.

*Positron-ring-plus-target scheme:*
The positron-ring-plus-target scheme could be tested at DAΦNE after the SIDDHARTA-2 run, perhaps starting as early as 2020 [59]. The goals of such a test would be two-fold: validation of the beam dynamics of the positron beam; and measurements of the thermo-mechanical stress in the target. These studies will be highly informative even though the beam energy will be significantly different from that required for muon production. A proposal for a primary electron-positron beam facility at CERN, based on a 3.5 GeV X-band linac, has recently been considered. Such a beam can be transferred in the SPS and accelerated up to 16 GeV where it could eventually be used for tests of the ring-plus-target scheme [54].

*High rate positron source*
R&D on a high rate positron source could take advantage of significant synergies with other future collider studies. The required positron source for the LEMMA scheme is challenging. The intensity required is two orders of magnitude higher than is specified for the LHeC ERL option [60] – this corresponds to a positron production rate of the order of $10^{18}$ Hz and $N_b(e^+) = 3\times10^{11}$ in the positron ring. These large values are synergistic with developments for state-of-the-art future colliders such as FCC-ee, ILC and CLIC.

The required intensity is strongly correlated with the beam lifetime that is determined by the ring momentum acceptance and the target material (see Sects. 2 and 3). The beam lifetime with the present optics is in the range of 40-50 turns corresponding to a positron loss rate of 2-3% per turn.

*Embedded positron source*
In the interaction of the primary positron beam with the beryllium target, bremsstrahlung photons are produced with a strong boost along the primary beam direction. It is possible to exploit the resulting photon flux to create an embedded positron source by placing a thick, high-Z target downstream of the muon target. This second target can then be used for electron positron pair production. Experimental tests of such a production scheme based on an adiabatic matching device have been performed at KEK [61, 62].

However, a system that is able to transform the temporal structure of the produced positrons to one compatible with the requirement of a standard positron injection chain is not yet available.

A full Monte Carlo simulation has been performed in order to evaluate the performance of such a scheme. In summary, it has been found that this scheme allows the production of roughly 60 secondary positrons from the interaction of 100 primary positrons, where <3 primary positrons are lost in the beryllium target. It would thus be sufficient to have a collection efficiency for these secondary positrons of ~5% in order to be able to make up for the losses in the primary beam.

R&D on the positron production target, where $e^+$ are produced by $e^+ e^-$ pair conversion of high energy gamma rays, would be useful in order to evaluate the thermo-mechanical stresses given by the deposited power and the integrated Peak Energy Deposition Density (PEDD).

**4.2. *Emittance Cooling***

As described in Sec. 2.1.2, the key technology R&D required for the demonstration of the next-generation cooling designs developed by MAP has been completed. In particular:
- the MAP design effort has provided cooling channel designs that, for the most part, meet or

exceed the specifications required for the HEP machines being considered;
- clear solutions exist for operating normal-conducting RF cavities in high magnetic fields;
- operational magnets now exceed the MAP design specifications for the highest fields required;
- the MICE experiment has successfully characterized the key parameters of the absorber materials on which cooling channel designs are based and has successfully measured the ionization cooling process.

Thus, there are two clear steps to take in moving forward. First, an iteration on the MAP designs to further optimize the cooling channel performance based on experimentally measured parameters would provide further confidence in our ability to achieve the performance parameters required for NF and MC applications. Second, construction of a prototype late-stage 6D cooling channel cell would pave the way to a characterization of its performance with beam.

A beam test of 6D cooling channel components would ideally utilize a source of pulsed muons in the correct momentum slice for detailed characterization of the cooling performance. We note that such a source was an integral part of the nuSTORM proposal, where the uncaptured beam at the end of the injection straight would be sent through an iron degrader. Simulations indicate that roughly $10^6$ muons/pulse in a suitable energy slice for cooling studies would be available downstream of the degrader. Given the recent tantalizing results from the MiniBoone collaboration [63] on the potential existence of sterile neutrinos, we can only note that pursuit of nuSTORM to fully study this sector would provide an ideal source of muons for a cooling demonstration. Of course, other potential sources must also be explored.

### 4.3. *Acceleration*

Muon beams must be accelerated to high energy in a very short period of time to avoid unacceptable decay losses. Since synchrotron radiation is not a limiting factor in accelerating muons to the TeV-scale, the efficiency of multi-pass acceleration makes it the preferred path to providing cost-effective collider facilities.

For acceleration from the ~100 GeV-scale to the TeV-scale, the MAP baseline utilized a hybrid RCS concept [36]. As noted in Sec. 2.1.2, the fast-ramping magnets required for such an accelerator must be capable of ±2 T peak-to-peak operation at a minimum frequency of 400 Hz. Because the MAP designs envisioned no more than a 15 Hz repetition rate for the accelerated muon beams, such a hybrid RCS scheme could readily accommodate injection during the rising half of the magnetic cycle and provide a highly efficient acceleration process. On the other hand, the LEMMA scheme utilizes a natural cycle time of ~2.2 kHz and cannot be matched to the slower ramp rate of the MAP hybrid RCS concept.

For LEMMA, alternative acceleration options that do not require ramped magnets must be explored. These options include RLA and FFAG [21] machines with large energy acceptance. An FFAG-based concept is presently being constructed for the CBETA project [64].

In each of the above cases, further magnet development along with detailed studies of the required lattices is highly desirable. Furthermore, customized concepts that consider re-use of the LHC tunnel as in Ref. [58, 65] are interesting and should also be studied in greater detail. Finally, acceleration options that leverage the smaller beam emittances available in the LEMMA scheme should be explored. In particular, acceleration technologies developed as part of the long-standing $e^+e^-$ linear collider research effort could potentially be of benefit.

### 4.4. *Collider*

The MAP effort produced a set of collider lattices with detailed interaction region (IR) optics [46]. These designs incorporated the necessary shielding constraints to deal with the impact of muon decay backgrounds on both the magnets and the detector. A design for a MC with the same level of maturity based on the LEMMA scheme has not yet been completed. In particular, optimization of the design may take advantage of the smaller bunch charges and lower emittance that are targeted. A particular challenge is the aggressive IR parameters that have been proposed. The design of the collider optics with the IR must be completed in order to assess the muon beam parameters and the luminosity that can ultimately be achieved. The LEMMA concept aims at $5\times10^{34}$ cm$^{-2}$s$^{-1}$ luminosity (see Table 4). Further studies are planned to determine the lower limit for the muon beam emittance, which is presently assumed to reach the ultimate goal of 40 nm.

The LEMMA final focus design with the nanobeam scheme will determine the feasibility of the very low $\beta^*$ being considered in order to leverage the low emittance. This design also allows a very small bunch length, which will require detailed study of the performance constraints imposed by collective effects and wake-fields constraints. These effects will have to be studied in complete detail. Finally, beam-beam simulations with this scheme will be necessary to determine the impact on the emittance as well as on the IR backgrounds induced in the detectors.

The bunch intensity in the collider for the LEMMA scheme relies on achieving bunch combination with no luminosity degradation from merging in either the longitudinal [54] or the transverse plane [55]. Detailed optics and beam dynamics studies will be essential to determine whether such performance can be achieved without significant emittance growth, thus providing the targeted luminosity performance.

### 4.5. *Detector*

Any detector at a MC will have to be designed with the backgrounds from muon decays fully accounted for. These issues have been studied in the context of the MAP effort [66]. Current approaches to handling high detector backgrounds appear adequate to preserving the required physics capabilities of the detector. These studies need to be extended, however, to study a full range of physics processes in detail in order to characterize the overall physics reach of any MC design.

### 5. Conclusion

Muon Colliders offer tremendous potential for the future of high energy physics research. The design studies and R&D conducted as part of the Muon Accelerator Program have considerably advanced the feasibility and our understanding of the anticipated performance of the proton-driver based muon generation scheme. In particular, the MAP R&D results on cooling channel design and technology demonstrations, along with the cooling measurements achieved by the MICE Collaboration, pave the way for a definitive demonstration of 6D cooling technology. Furthermore, the design studies undertaken by the MAP team provide a basis for providing a range of physics capabilities, from Neutrino Factories to Muon Colliders, depending on the high energy physics community's identified scientific needs.

The LEMMA scheme offers an attractive route to a low emittance muon beam produced by a positron driven muon source. Such a source, which would not require muon cooling, may allow operation of a very high energy Muon Collider with manageable neutrino radiation on and off the site. This scheme has been sketched with a proposed parameter table giving its energy reach and possible luminosity. Major R&D topics to address key issues of the various technologies and demonstrate their feasibility have been outlined.

Finally, the recent consideration of a 14 TeV Muon Collider in the LHC tunnel, which would take advantage of the existing accelerator infrastructure and injector systems at CERN, could probe the same energy scale as a 100 TeV proton-proton collider, but at lower cost and with cleaner physics conditions in the detector. Such an approach, which could be based on either type of muon source, is especially attractive. Nonetheless, further detailed study is required.

## 6. Acknowledgments

We would like to thank the MAP Collaboration, the MICE Collaboration, the International Design Study for a Neutrino Factory, and the LEMMA Collaboration for providing the information for this review article and also for their dedicated efforts to carry forward the development of these challenging muon accelerator concepts that have tremendous potential for high energy physics research.

This work was supported, in part, by the U.S. Department of Energy under contract DE-SC0012704.